
\input amstex
\documentstyle{amsppt}
\document
\NoRunningHeads
\TagsAsMath
\TagsOnRight
\NoBlackBoxes

\define\DET{\text{det}}

\define\laaa#1{\lambda_1 ,\cdots , \lambda_{#1}}
\define\gaaa#1{\gamma_1,\cdots,\gamma_{#1}}
\define\teet#1#2{\theta [\eta _{#1}] (#2)}
\define\tede#1{\theta [\delta](#1)}
\define\taaa #1{\tau _1,\cdots, \tau _{#1}}
\topmatter
\title
A New Set of Exact Form Factors.
\endtitle
\font\special=cmr6
\author
Feodor A. Smirnov\\
{\special Research Institute for Mathematical Sciences,
Kyoto  University,}\\{\special Sakyo-ku 606, Kyoto, JAPAN}\\
{\special and}\\
{\special St. Petersburg Branch of Steklov Mathematical Institute,}\\
{\special Fontanka 27, St. Petersburg 191011,  RUSSIA}
\endauthor
\abstract
We present form factors for
a wide list of integrable models which includes
marginal perturbations of $SU(2)$ WZNZ model
for arbitrary central charge and the principal chiral field model.
The interesting structure of these form factors is discussed.
\endabstract
\endtopmatter

\head
1. Introduction
\endhead
The form factor bootstrap is powerful method of study of
integrable field theories. There are many models for which the
form factors are known, the most important examples are
Sine-Gordon model, its asymptotically free limit: $SU(2)$-
invariant Thirring model, $O(3)$ - nonlinear sigma model,
$SU(N)$-invariant Thirring model [1].
This list seems to be representative enough. However,
for better understanding of integrable field
theory we still need some more examples. Asymptotically
free theories are of primary importance for further investigations
in the field. They allow important type of quantum symmetry:
Yangian symmetry [2], which is in less modern language the
same as L\"usher nonlocal charges with interesting
classical limit [3]. The knowledge of
exact out of shell solutions of asymptotically free integrable theories
should allow to understand how do the usual tools of
modern theoretical physics (functional integration)
apply to them.  We consider this problem as the
most important one: we should learn from exact solutions
how to perform functional integration.
There are also other reasons why such models as perturbations
of WZNW model and principal chiral field are interesting for
the application of form factor bootstrap which will be clear later.

\def\b{\beta}
Suppose we deal with an massive integrable model
which contains only one particle in the spectrum
(with isotopic degrees of freedom)
with the two-particle
S-matrix $S(\b)$ which satisfies Yang-Baxter, crossing, unitarity [4].
Then for the operator to be local it is necessary and sufficient
that its form factors (matrix element between vacuum and $n$-particle
state)
satisfy the system of equations [1] which naturally splits into two parts:
Riemann-Hilbert problem
$$\align
& f(\b _1,\cdots,\b _i,\b _{i+1},\cdots ,\b _{n})S(\b _i-\b _{i+1})=
f(\b _1,\cdots,\b _{i+1},\b _i, \cdots ,\b _{n}),\\
& f(\b _1,\cdots ,\b _{n-1},\b _{n}+2\pi i)=
f(\b _n,\b _1,\cdots,\b _{n-1}), \tag {1} \endalign
$$
and residue condition
$$\align
&2\pi i \ \text{res}_{\b _n=\b _{n-1}+\pi i}
\ f(\b _1,\cdots,\b _{n-2} ,\b _{n-1},\b _{n}) = \tag {2}\\ &=
f(\b _1,\cdots,\b _{n-2})\otimes s_{n-1,n}\bigl(I-S(\b _{n-1}-\b_1)
\cdots S(\b _{n-1}-\b _{n-2})\bigr)
\endalign $$
where the usual [1] conventions are made:
$f$ belongs to the tensor product of spaces $h^*$ related to particles
($h$ is one particle isotopic space),
$S(\b _i-\b _j)$ acts nontrivially only in the tensor product of spaces
related to particles with rapidities $\b _i ,\ \b _j$ ,
permutation of spaces is
applied if corresponding rapidities are permuted,
$s_{n-1,n}$ is a vector in the tensor product of
$(n-1)$-th and $n$-th spaces constructed from charge conjugation matrix,
$f$ is meromorphic function of all its arguments in finite part of
complex plane, as function of $\b _n$ it does not have other singularities
for $0\le \text{Im}\b _n \le 2\pi i$ but simple poles at the points
$\b _n=\b _j+\pi i, \ j\le n-1$.

Several solutions to this
infinite system of equations are known for the SG model
which correspond to the most important local operators. We shall
consider SG only in repulsive case when two component soliton
is the only particle in the spectrum, so the space $h$ is two-dimensional,
and $4\times 4$ S-matrix $S^{SG}_{\xi}(\b)$ depends on the
coupling constant $\xi$ : $\pi \le \xi \le \infty$. The explicit formula
for this S-matrix will be given later.
When ${\xi \over \pi}$
is rational the restriction in the space of states is possible [5]
which provides different model ($\phi _{1,3}$-perturbation of
minimal model [6]).
These restrictions do not respect the Hermitian structure of
the space of states of SG and
allow their own definite Hermitian structure for
${\xi \over \pi} $ integer. The S-matrix for the restricted models
is obtained from SG one by RSOS [7] procedure.
Generally, the idea of using RSOS restrictions for constructing physical
S-matrices is due to [8].
In this paper
we shall not go into much
details of the restrictions because the
restrictions of form factors occur quite naturally
and do not present much difficulty as far as SG form factors are known.
It should be mentioned also that for $\xi=\infty$ the SG S-matrix
produces the S-matrix of $SU(2)$-invariant Thirring model:
$$ S^{SG}_{\infty}(\b) =S^{ITM}(\b)$$

There are several models for which the one-particle isotopic
spaces are tensor product of two SG isotopic spaces (or their
restrictions)
the S-matrices are
different particular cases (and restrictions) of the
following one
$$ - S^{SG}_{\xi _1}(\b)\otimes S^{SG}_{\xi _2}(\b) \tag {3}$$
for two different coupling constants [9,10,11,12]. The most interesting
examples
are the following. For  $\xi _1=\infty,  \xi _2 =k+2$ after
restriction of the second S-matrix we deal with the perturbation
of WZNW-model on level $k$ with action
$$ S=S_{WZNW_k}+\lambda\int d^2 x J^a\bar{J}^a $$
which is the same as $k$-flavour $SU(2)$-Thirring model
due to Polyakov-Wiegmann bosonisation [9]. In extreme case
$\xi _1=\infty,  \xi _2 =\infty $ we get principal chiral field model (PCF)
[9,12].
These are the models we are mostly interested in. To get the
form factors for them we shall consider the most general S-matrix
of the type given by (3).

Trying to solve the equations (1,2) for the S-matrix (3) we
find the following amusing circumstance. Consider the Riemann-Hilbert
problem (1) for the S-matrix (3). Evidently, it is
satisfied by a slight
modification of the tensor product of SG form factors:
$$\text{exp}\bigl({1\over 2}\sum\b _j\bigr)
\prod _{i<j} \text{cth}{1\over 2}(\b _i-\b _j)
f_{\xi _1}(\b _1,\cdots,\b _n)\otimes f_{\xi _2}(\b _1,\cdots,\b _n) \tag {4}$$
Also this function is meromorphic and does not have other singularities
as function of $\b _n$ but usual simple poles at $\b _n=\b _j+\pi i$
(SG form factors vanish when $\b _i=\b _j $).
However, this anzatz breaks the equation (2).
So, we have to look for something more intelligent.
The lesson we learn from the naive anzatz (4) is that
if we consider any solutions to the Riemann-Hilbert
problem (1) for SG S-matrices then their tensor product satisfies
the same equations for the tensor product S-matrix, but
we should take care of the third equation.

On the other hand the Riemann-Hilbert problem (1) for SG model
can be considered as deformation of
Knizhnik-Zamolodchikov [13] (KZ) equations on level zero
for the algebra $U_q (\widehat{sl} (2))$ with
$q=\text{exp} ({2\pi ^2 i\over \xi})$. This is rather
informal way of thinking because strictly speaking
the deformations of
KZ equations (thorough vertex operators, highest weight
representations etc)
are properly defined for $q<1$ [14],
for Yangian case $q\to 1$ they can be treated in terms of
asymptotic series [15], for $|q|=1$ we still do not know how
to treat them, but fortunately on level zero we do know how to solve them!

The same equations for the S-matrix (3) can
be called deformed KZ for non semi-simple algebra
$U_{q_1} (\widehat{sl} (2))\otimes U_{q_2} (\widehat{sl} (2))$ with
$q_n=\text{exp} ({2\pi ^2 i\over \xi _n})$.
Certainly the solutions for this case should be given
by linear combinations of tensor products of different
solutions of equations for
$U_{q_1} (\widehat{sl} (2))$ and $U_{q_2} (\widehat{sl} (2))$.
So, our goal will be achieved if we know enough solutions
in these two cases in order to construct linear combination of
their tensor products which satisfy the residue condition (2).
As it had been mentioned in [1] and explained in
details for Yangian case in [15,16] we do know many solutions to
the Riemann-Hilbert problem (1)
in SG case (actually the same number as for usual KZ
equations),
but only very special ones were used for SG form factors because
we had to satisfy the residue equation. In this paper we shall show that
all the solutions are needed in order to construct the
form factors for the S-matrix (3) through the procedure explained above.

\head
2. Solution of Riemann-Hilbert problem for $U_{q} (\widehat{sl} (2)),\  |q|=1$.
\endhead

\def\f{\tilde{f}}
In this section we shall be interested in the solutions to
the Riemann-Hilbert problem (1). It is convenient to
change the sign in the RHS of the second equation, this
is harmless because the solutions to modified in this way
equations can be transformed into the solution of original ones via
multiplication by $\text{exp}\bigl({1\over 2}\sum\b _j\bigr)$. So, we
want to solve the equations:
$$\align
& \f(\b _1,\cdots,\b _i,\b _{i+1},\cdots ,\b _{2n})S_{\xi}(\b _i-\b _{i+1})=
\f(\b _1,\cdots,\b _{i+1},\b _i, \cdots ,\b _{2n}),\\
& \f(\b _1,\cdots ,\b _{2n-1},\b _{2n}+2\pi i)=
-\f(\b _{2n},\b _1,\cdots,\b _{2n-1}) \tag {5}\endalign $$
where the SG S-matrix is given by [4]
$$\align &S _{\xi}(\b)=
\frac {S_{\xi,0}(\b)} { \text{sh}{\pi \over \xi}(\b -\pi i)}
\widehat{S}_{\xi}(\b), \qquad
S_{\xi,0}(\b)=-\text{exp}\bigl(-i\int\limits _0^{\infty}
\frac {\text{sin}(k\b)\text{sh}({\pi-\xi \over 2}k) }
{k\text{sh}({\pi k\over 2})\text{ch}({\xi k\over 2}) }
\bigr),\\
&\widehat{S}_{\xi}(\b)  =
\frac {S_{\xi,0}(\b)} { \text{sh}{\pi \over \xi}(\b -\pi i)}
\pmatrix
 &\text{sh}{\pi \over \xi}(\b -\pi i),&0 &0                              &0\\
 &0    & -\text{sh}{\pi \over \xi}\b    &\text{sh}{\pi^2 i \over \xi}     &0\\
 &0    & \text{sh}{\pi^2 i\over \xi}    &-\text{sh}{\pi \over \xi}\b    &0\\
 &0 &0                             &0   &\text{sh}{\pi \over \xi}(\b -\pi i)
\endpmatrix ,\\
\endalign $$
Later we shall omit index $\xi$ when only one $U_q(\widehat{sl}(2))$
is involved.
We shall consider the solution to these equations of special
isotopic character. The algebra $U_{q} (\widehat{sl} (2))$ contains two finite
dimensional subalgebras isomorphic to
$U_{q} (sl (2)) $. The isotopic spaces of particles can be
considered as the spaces of two-dimensional representations
for these two subalgebras
(the explicit formulae can be found in [5]).
We shall restrict ourselves with the
consideration of those solutions which belong to
invariant with respect to one of these subgroups spaces in the
tensor product (the solutions will carry an index $\gamma=\pm$ to
indicate that they are invariant with respect to one or another
subalgebra). For SG case these solutions describe
the most fundamental local operators: energy-momentum tensor
and current.
This consideration can be generalized
for $|q|<1$ case [17].

Let us mention that the solutions to the equations (5) are
defined up to multiplication by arbitrary quasiconstant:
$2\pi i$-periodic, symmetric function of $\b _j$.
The solutions which will be presented later are supposed to
constitute the full set
of meromorphic solutions of given
isotopic structure up to quasiconstants.

\def\a{\alpha}
For what follows we shall need two special functions: $\varphi(\b)$
and $\zeta (\b)$. We shall not write down explicit formulae for
these functions which can be found in [1], but just present their
most important properties:
$$\align
&  \varphi(\b-2\pi i)=  \varphi(\b) \frac
{\text{sh}{\pi \over \xi}(\b -{\pi i\over 2})}
{\text{sh}{\pi \over \xi}(\b -{3\pi i\over 2})},
\qquad \varphi(\b-\pi i)\varphi(\b)= \frac {1}
{2\text{sh}{\pi \over \xi}(\b -{\pi i\over 2})
\text{sh}(\b -{\pi i\over 2})}\\
&\frac{\varphi(\b+{\pi i\over 2})}{ \varphi(\b-{\pi i\over 2})}=
S_{0}(\b),\qquad
\zeta(\b) S_{0}(\b) = \zeta(-\b),\\& \zeta(\b-2\pi i) = \zeta(-\b),\qquad
\zeta(\b)\zeta(\b-\pi i)=\bigl(\varphi(\b+{\pi i\over 2})\bigr)^{-1}
\tag {6}
\endalign $$
We shall also need the vector-functions $\frak{F}^{\gamma}_n
(\a _1,\cdots,\a _{n-1}|\b_1,\cdots,\b_{2n})$
defined by the following list of requirements:
\item {1)}
they take values in the tensor product of $2n$ spaces $h^*$ ($h\sim \Bbb C
^2$),
\item {2)}
they are antisymmetric, entire, periodic with period $2\xi i$ functions
of $\a _1,\cdots,\a _{n-1}$
\item {3)} they are antiperiodic with period $2\xi i$ functions
of $\b_1,\cdots,\b_{2n}$ satisfying the following symmetry property:
$$ \align &\frak{F}^{\gamma}_n (\a _1,\cdots,a _{n-1}|\b_1,\cdots,\b _i,\b
_{i+1},\cdots,\b_{2n}) \widehat{S}_{\xi}(\b _i-\b _{i+1})=\\&=
\frak{F}^{\gamma}_n (\a _1,\cdots,\a _{n-1}|\b_1,\cdots,\b _{i+1},\b _i,
\cdots,\b_{2n}), \endalign $$
\item {4)} for $\gamma=+$ or $-$ they are singlets with respect to one or
another finite-dimensional quantum group as explained above,
\item {5)}
they do not have other singularities but simple poles at the points $\b _j=\b
_i+\pi i +k \xi i$ for $j>i,\ k\in\Bbb Z$
\item {6)}
they satisfy the recurrent relations:  $$\align
&\text{res} _{\b_{2n}=\b _{2n-1}+\pi i}\qquad\frak{F}^{\gamma}_n
(\a _1,\cdots,\a _{n-1}|\b_1,\cdots,\b _{2n-2}, \b _{2n-1},\b_{2n})=\\& =
{\xi \over \pi} (e_{2n-1,-}\otimes e_{2n,+}+e_{2n-1,+}\otimes e_{2n,-})
\otimes \\&
\bigl( \sum\limits _{l=1}^{n-1} (-1)^l
\frak{F}^{\gamma}_{n-1}
(\a _1,\cdots,\widehat{\a _l},\cdots,\a _{n-1}|\b_1,\cdots,\b _{2n-2})
\ \text{exp}\bigl({\pi\over\xi}(n-1)(\a _l-\b _{2n-1})\bigr)\\
&\times\bigl[ \prod\limits _{p=1}^{2n-2}
\text{sh}{\pi\over \xi}(\a _l-\b _p -{\pi i\over 2})
-q^{(n-1)}\prod\limits _{p=1}^{2n-2}
\text{sh}{\pi\over \xi}(\a _l-\b _p -{\pi i\over 2})
\bigr] \bigr)
\\ &\times
\prod\limits _{q=1}^{n-1}
\text{sh}{\pi\over \xi}(\a _q-\b _{2n-1} -{\pi i\over 2})
\bigl(\prod\limits _{i=1}^{2n-2}\text{sh}{\pi\over\xi}(\b_{2n-1}-\b_i)
\bigr)^{-1}\tag {7}
\endalign $$
where $e_{i,\pm}$ is basis in $i$-th space, the vector
$$e_{2n-1,-}\otimes e_{2n,+}+e_{2n-1,+}\otimes e_{2n,-} \tag {8}$$
is nothing but
$s_{2n-1,2n}$ from (2).
\item {7)}
for n=1 we have
$$
\align
&\frak{F}^{\gamma}_ 1(\b_1,\b_{2})= \\&
\frac{\text{exp}(\gamma{\pi\over 2\xi}
(\b _2-\b _1-\pi i)) \bigl(e_{1,+}\otimes e_{2,-}\bigr) +
\text{exp}(-\gamma{\pi\over 2\xi}
(\b _2-\b _1-\pi i)) \bigl(e_{1,-}\otimes e_{2,+}\bigr) }
{ \text{sh}{\pi \over \xi }(\b _2-\b _1-\pi i)} \endalign $$

It is possible to give explicit formulae for these
functions in terms of certain determinants [1], we shall
need it only in the last two sections of this paper.

Now we are ready to write down the solutions.
Different solution will be counted
by $\gamma =\pm$ which specifies
the isotopic structure (singlet with respect
to one or another finite-dimensional quantum subalgebra) and
a set of integers $\{k_1,\cdots,k_{n-1}\}$ such that
$|k_i|<n-1, \ \forall i$.
The sets $\{k_1,\cdots,k_{n-1}\}$ play the same role as different
contours
in integral formulae for solutions of
KZ equations [13,18,19]. The solutions are given by
$$\align
& \f^{\gamma}_{k_1,\cdots,k_{n-1}} (\b_1,\cdots,\b _{2n})=
d^n\ \prod\limits_{i<j}\zeta(\b _i-\b _j) \tag {9}\\ &
\int\limits _{-\infty}^ {+\infty}d\a _1\cdots
\int\limits _{-\infty}^ {+\infty}d\a _{n-1}\prod\limits _{i=1}^{n-1}
\prod\limits _{j=1}^{2n}\varphi(\a _i-\b _j)\text{exp}\bigl(
\sum\limits _{i=1}^{n-1}\a _i k_i \bigr)
\frak{F}^{\gamma}_n
(\a _1,\cdots,\a _{n-1}|\b_1,\cdots,\b_{2n}) \endalign $$
where
$$d={1\over 4\pi\xi i \zeta (\pi i)}$$
The integrals in (9) are regularized in a special way explained in [1],
this regularization makes sense only for $k_i$ integer,
the limitation $|k_i|<n-1$ comes from requirement of convergency
of regularized integrals. As it is shown in [1] the functions
$\f^{\gamma}_{k_1,\cdots,k_{n-1}} (\b_1,\cdots,\b _{2n})$ do
satisfy (5).

Let us make some comments on the sets  $\{k_1,\cdots,k_{n-1}\}$
which count the solutions. First, due to antisymmetry of
$\frak{F}^{\gamma}_n $
with respect to $\a$'s we must consider only those sets with
all $k_i$ different. Second, for "good" entire, $2\xi i$-periodic
function  $F(\a)$ we have the following relation:
$$\sum\limits_{i=1}^{n}\sigma _{2i-1}(e^{\b _1},\cdots,
e^{\b _{2n}})
\int\limits _{-\infty}^ {+\infty}d\a
\prod\limits _{j=1}^{2n}\varphi(\a-\b _j)
\text{exp}((n+1-2i)\a) F(\a) =0 \tag {10}$$
where $\sigma _k$ is elementary symmetrical polynomial of degree $k$.
It is explained in [1,16] what kind of functions is "good", we would not
go into details here, at least the function  $\frak{F}^{\gamma}_n$
considered as a function of one $\a$ is "good". The formula (10) gives
one (and only one) relation of
linear dependence with quasiconstant coefficients
between the integrals with different $k$. It provides, certainly,
linear dependence between some solutions to (5), to get
really different solutions we could, for example, require $k_i\ne n-1
\ \forall i$.

There is a beautiful way of understanding the solutions (9)
explained in details (for Yangian case) in [15,16].
The point is that the solutions of KZ equations for
the case of $sl(2)$, on level zero are given by determinants
composed of
periods of certain second kind differentials
on hyperelliptic surface. The size of these determinants
is equal to genus of the surface while the number of different
cycles in twice bigger, that gives rise to different
solutions counted by different subsets of cycles.
The deformation of this picture should be understood as
follows. The points $\b _1,\cdots\b _{2n}$ are
branching points of "quantum hyperelliptic surface",
the integral
$$\text{exp}(-{1\over 2}\sum \b _p)\sigma _{2i-1}(e^{\b _1},\cdots,
e^{\b _{2n}})
\int\limits _{-\infty}^ {+\infty}d\a
\prod\limits _{j=1}^{2n}\varphi(\a-\b _j)
\text{exp}(k\a) F(\a)  $$
is the period of differential defined by
$2\xi i$-periodic function $ F(\a)$
taken over the cycle around two branching points:
$\b _{n-k},\ \b _{n-k+1}$. The "good" functions mentioned
above are analogues of those differentials which do
not have simple poles on surface.
If we consider first the Yangian limit ($\xi\to\infty$)
and then classical limit of Yangian then the correspondence can
be explicitly understood in terms of asymptotics [15,16,20].

\head
3. Form factors for SG model and its restrictions.
\endhead

Let us return to SG model. The form factors should satisfy not
only the Riemann-Hilbert problem (1), but also the residue condition
$$\align
&2\pi i \ \text{res}_{\b _{2n}=\b _{2n-1}+\pi i}
\ f(\b _1,\cdots,\b _{2n-2} ,\b _{2n-1},\b _{2n}) = \\ &=
f(\b _1,\cdots,\b _{2n-2})\otimes s_{2n-1,2n}\bigl(I-S(\b _{2n-1}-\b_1)
\cdots S(\b _{2n-1}-\b _{2n-2})\bigr)
\endalign $$
So, we have to consider the problem of calculation of residues
of the kind.

It is easy to show that generally the expressions (9)
have simple poles at $\b _{2n}=\b _{2n-1}+\pi i $.
The recurrent relations (7) are useful for the calculation
of residues because together with the properties of
$\varphi$ (6) they provide that the
integral with respect to $\a _l$ in $l$-th term of (7)
substituted to (9) can be replaced by contour integral
over the contour: $(-\infty ,\infty)$ $(\infty ,\infty +2\pi i)$
$(\infty +2\pi i,-\infty+2\pi i)$ $(-\infty +2\pi i ,-\infty)$.
So, this integral is calculated via the poles of the integrand
which are situated at the points $\a _l=\b _{2n-1}+{\pi i\over 2}$,
$\a _l=\b _{2n-1}+{3\pi i\over 2}$. However, generally the
expressions for the residues can not
be expressed in terms of functions of the type (9) with $2n-2$
points $\b _j$. There are only three possibilities to combine
the functions (9) with different $\{k _i\}$ in order to
have nice expressions:
$$\align
& 2\pi i \ \text{res}_{\b _{2n}=\b _{2n-1}+\pi i}
\ d^n\ \prod\limits_{i<j}\zeta(\b _i-\b _j)
\int\limits _{-\infty}^ {+\infty}d\a _1\cdots
\int\limits _{-\infty}^ {+\infty}d\a _{n-1}\prod\limits _{i=1}^{n-1}
\prod\limits _{j=1}^{2n}\varphi(\a _i-\b _j)\\ & \times
\prod\limits _{i=1}^{n-1}2\text{ch}(\a _i- \b _{2n-1})
\text{exp}\bigl(\sum\limits _{i=1}^{n-1}\a _i l_i \bigr)
\frak{F}^{\gamma}_n
(\a _1,\cdots,\a _{n-1}|\b_1,\cdots,\b_{2n})=0,\\&{ }\\
& 2\pi i \ \text{res}_{\b _{2n}=\b _{2n-1}+\pi i}
\ d^n\ \prod\limits_{i<j}\zeta(\b _i-\b _j)
\int\limits _{-\infty}^ {+\infty}d\a _1\cdots
\int\limits _{-\infty}^ {+\infty}d\a _{n-1}\prod\limits _{i=1}^{n-1}
\prod\limits _{j=1}^{2n}\varphi(\a _i-\b _j)\\ & \times
\bigl(1+ie^{\a _1-\b_{2n-1}}\bigr)
\prod\limits _{i=2}^{n-1}2\text{ch}(\a _i- \b _{2n-1})
\text{exp}\bigl(\sum\limits _{i=1}^{n-1}\a _i l_i \bigr)  \tag {11}
\\&\times
\frak{F}^{\gamma}_n
(\a _1,\cdots,\a _{n-1}|\b_1,\cdots,\b_{2n})=2
s_{2n-1,2n}\otimes \f^{\gamma}_{l_1,\cdots,l_{n-1}} (\b_1,\cdots,\b _{2n}),\\
&{ }\\
& 2\pi i \ \text{res}_{\b _{2n}=\b _{2n-1}+\pi i}
\ d^n\ \prod\limits_{i<j}\zeta(\b _i-\b _j)
\int\limits _{-\infty}^ {+\infty}d\a _1\cdots
\int\limits _{-\infty}^ {+\infty}d\a _{n-1}\prod\limits _{i=1}^{n-1}
\prod\limits _{j=1}^{2n}\varphi(\a _i-\b _j)\\ & \times
\bigl(1-ie^{\a _1-\b_{2n-1}}\bigr)
\prod\limits _{i=2}^{n-1}2\text{ch}(\a _i- \b _{2n-1})
\text{exp}\bigl(\sum\limits _{i=1}^{n-1}\a _i l_i \bigr)
\\&\times
\frak{F}^{\gamma}_n
(\a _1,\cdots,\a _{n-1}|\b_1,\cdots,\b_{2n})= \\&= -2
s_{2n-1,2n}\otimes \f^{\gamma}_{l_1,\cdots,l_{n-1}} (\b_1,\cdots,\b _{2n})
S(\b _{2n-1}-\b_1)
\cdots S(\b _{2n-1}-\b _{2n-2}),
\endalign $$
The limitations for the values of $l_i$ are clear in every case.

The solutions for the complete system of relations (1,2) for $n>1$
are given by
$$  \align
&f^{\gamma}_{+} (\b_1,\cdots,\b _{2n})=
\text{exp}({1\over 2}\sum\b _j)
\f^{\gamma}_{-(n-1),-(n-3),\cdots,(n-5),(n-3)}
(\b_1,\cdots,\b _{2n}),\\
&f^{\gamma}_{+} (\b_1,\cdots,\b _{2n})=
\text{exp}({1\over 2}\sum\b _j)
\f^{\gamma}_{-(n-3),-(n-5),\cdots,(n-3),(n-1)}
(\b_1,\cdots,\b _{2n})      \tag {12}
\endalign $$
To prove that these functions do satisfy the residue
condition we notice that due to antisymmetry of
$\frak{F}^{\gamma}_n$ under the integral in (9)
the expression
$$\text{exp}\bigl(\sum\limits _{i=1}^{n-1}
(n+1-2i)\a _i\bigr) $$
can be replaced by
$$\align
&{1\over 2}\text{exp}((n-2)\b _{2n-1}-(n-3)\a _1)
\left[\bigl(1+ie^{\a _1-\b_{2n-1}}\bigr)+
\bigl(1-ie^{\a _1-\b_{2n-1}}\bigr)\right]\\ &\times
\prod\limits _{i=2}^{n-1}2\text{ch}(\a _i- \b _{2n-1})
\text{exp}\bigl(\sum\limits _{i=2}^{n-1}
(n-2i)\a _i\bigr)
\endalign $$
Now one just uses (11) to prove that the residue condition
is satisfied.

The functions (12) are not independent, due to (10) they
satisfy the relation
$$ \bigl(\sum e^{-\beta _j} \bigr)
f^{\gamma}_{+} (\b_1,\cdots,\b _{2n})=
\bigl(\sum e^{+\beta _j} \bigr)
f^{\gamma}_{-} (\b_1,\cdots,\b _{2n}) $$

Analyzing the two particle form factor
one realizes that the form factors of energy-momentum
tensor $T_{\sigma _1 \sigma _2}$ and $U(1)$ current
$J_{\sigma}$ (we use light-cone components,
$\sigma =\pm$) are given by [1]:
$$\align
&f_{++} (\b_1,\cdots,\b _{2n})=
\bigl(\sum\limits_{j=1}^{2n} e^{\b _j}\bigr)\sum _{\gamma}
f^{\gamma }_{+} (\b_1,\cdots,\b _{2n}),\\
&f_{--} (\b_1,\cdots,\b _{2n})=
\bigl(\sum\limits_{j=1}^{2n} e^{-\b _j}\bigr)\sum _{\gamma}
f^{\gamma}_{-} (\b_1,\cdots,\b _{2n}),\\
&f_{+-} (\b_1,\cdots,\b _{2n})=\\
&=\bigl(\sum\limits_{j=1}^{2n} e^{-\b _j}\bigr)\sum _{\gamma }
f^{\gamma }_{+} (\b_1,\cdots,\b _{2n})=
\bigl(\sum\limits_{j=1}^{2n} e^{\b _j}\bigr)\sum _{\gamma }
f^{\gamma }_{-} (\b_1,\cdots,\b _{2n}),\\
&f_{\pm} (\b_1,\cdots,\b _{2n})=
\sum _{\gamma }
(-)^{\gamma } f^{\gamma}_{\pm}
(\b_1,\cdots,\b _{2n})    \tag {13}
\endalign $$

There is one more set of solutions which have
good residues:
$$ \f^{\gamma}_{-(n-2),-(n-4),\cdots,(n-4),(n-2)}
(\b_1,\cdots,\b _{2n}) \tag {14}$$
but in that case we keep minus in the RHS of (1), also we have plus
instead of minus between two terms in the RHS of (2), hence
these solutions
corresponds to disorder
operators [1].

Let us explain briefly the restrictions of SG.
Consider the modified energy-momentum tensor whose form factors
are given by
$$\align
&f_{++}^+ (\b_1,\cdots,\b _{2n})=
\bigl(\sum\limits_{j=1}^{2n} e^{\b _j}\bigr)
f^{+}_{+} (\b_1,\cdots,\b _{2n}),\\
&f_{--}^+ (\b_1,\cdots,\b _{2n})=
\bigl(\sum\limits_{j=1}^{2n} e^{-\b _j}\bigr)
f^{+}_{-} (\b_1,\cdots,\b _{2n}),\\
&f_{+-}^+ (\b_1,\cdots,\b _{2n})=\\&=
\bigl(\sum\limits_{j=1}^{2n} e^{-\b _j}\bigr)
f^{+}_{+} (\b_1,\cdots,\b _{2n})=
\bigl(\sum\limits_{j=1}^{2n} e^{\b _j}\bigr)
f^{+}_{-} (\b_1,\cdots,\b _{2n})
\endalign $$
These operators is invariant with respect to one of $U_q(sl(2))$
subalgebras. That is why for rational $\xi$ the intermediate
states in the correlations of these operators among themselves
happen to be truncated [5]. So, we can construct a new restricted
theory (RSG) with smaller operator content and truncated
space of states. This theory is well known to coincide
with $\phi _{1,3}$-perturbation of minimal model [6].
The truncation does not respect the
Hermitian structure of SG space of states, so, to have
RSG to be defined intrinsically we have to introduce new
Hermitian structure. It can be made positively defined
in the region on coupling constant we deal with only
for ${\xi\over \pi}$ integer
and ${\xi\over \pi}={3\over 2}$ (in what follows
we shall not be interested in the
latter case).
One of disorder operators allows
restriction, it coincides with the operator $\phi _{1,2}$ in
ultraviolet limit.

Thus, only very small part of solutions we know is useful for SG form factors.
However, in the next section we shall show that all of them are needed
for the models with the tensor product S-matrices.

\head
4. The form factors for tensor product S-matrix.
\endhead

The Riemann-Hilbert problem (1) looks in that case as
$$\align
& f(\b _1,\cdots,\b _i,\b _{i+1},\cdots ,\b _{2n})
S_{\xi _1 \xi _2}(\b _i-\b _{i+1})=
f(\b _1,\cdots,\b _{i+1},\b _i, \cdots ,\b _{2n}),\\
& f(\b _1,\cdots ,\b _{2n-1},\b _{2n}+2\pi i)=
f(\b _{2n},\b _1,\cdots,\b _{2n-1}) \endalign$$
with
$$ S_{\xi _1 \xi _2}(\b)=
- S^{SG}_{\xi _1}(\b)\otimes S^{SG}_{\xi _2}(\b) \tag {15}$$
Evidently, these two equations are satisfied by any expression of the
kind:
$$\text{exp}\bigl(\pm{1\over 2}\sum\b _j\bigr)
\prod _{i<j} \text{cth}{1\over 2}(\b _i-\b _j)
\f^{\gamma _1}_{\xi _1,\ K} (\b_1,\cdots,\b _{2n})\otimes
\f^{\gamma _2}_{\xi _2,\ L} (\b_1,\cdots,\b _{2n}) $$
where we denoted the ordered subsets of integers
$\{k_1,\cdots,k_{n-1}\}:\ |k_i|\le n-1 $ and
$\{l_1,\cdots,l_{n-1}\}:\ |l_i|\le n-1 $ by $K$ and $L$.
The problem is to satisfy the residue condition:
$$\align
&2\pi i \ \text{res}_{\b _{2n}=\b _{2n-1}+\pi i}
\ f(\b _1,\cdots,\b _{2n-2} ,\b _{2n-1},\b _{2n}) = \\ &=
f(\b _1,\cdots,\b _{2n-2})\otimes \tilde{s}_{2n-1,2n}\bigl(I-
S_{\xi _1 \xi _2}(\b _{2n-1}-\b_1)
\cdots S_{\xi _1 \xi _2}(\b _{2n-1}-\b _{2n-2})\bigr) \tag {16}
\endalign $$
where $\tilde{s}_{2n-1,2n}$ is the tensor product of two
vectors like (8).
We shall show that these equations are satisfied by
$$ \align
&f^{\gamma _1\gamma _2}_{\pm\ \xi _1\xi _2} (\b_1,\cdots,\b _{2n})= (2\pi)^n
\text{exp}\bigl(\pm{1\over 2}\sum\b _j\bigr)
\prod _{i<j} \text{cth}{1\over 2}(\b _i-\b _j)
\\&\times \sum\limits _{N_{\pm}=K\cup L}
\f^{\gamma _1}_{\xi _1,\ K} (\b_1,\cdots,\b _{2n})\otimes
\f^{\gamma _2}_{\xi _2,\ L} (\b_1,\cdots,\b _{2n})    \tag {17}
\endalign $$
where $N_ +=\{-(n-1),-(n-2),\cdots ,(n-3), (n-2)\}$ ,
$N_ -=\{-(n-2),-(n-3),\cdots ,(n-2), (n-1)\}$.
To prove that we have to calculate the residue. It is not
complicated, first let us write more explicit formulae:
\def\aa{\tilde{\alpha}}
$$\align
&f^{\gamma _1\gamma _2}_{\pm\ \xi _1\xi _2} (\b_1,\cdots,\b _{2n})=
{1\over ((n-1)!)^2}d_{\xi _1 \xi _2}^n
\text{exp}\bigl(\pm{1\over 2}\sum\b _j\bigr)
\prod _{i<j} {\zeta}_{\xi _1\xi _2}(\b _i-\b _j) \times\\
&
\int\limits _{-\infty}^ {+\infty}d\a _1\cdots
\int\limits _{-\infty}^ {+\infty}d\a _{n-1}
\int\limits _{-\infty}^ {+\infty}d \aa _1\cdots
\int\limits _{-\infty}^ {+\infty}d\aa _{n-1}
\prod\limits _{i=1}^{n-1}
\prod\limits _{j=1}^{2n}\varphi _{\xi _1}(\a _i-\b _j)
\prod\limits _{i=1}^{n-1}
\prod\limits _{j=1}^{2n}\varphi _{\xi _2}(\aa _i-\b _j)
\\&\times
\text{exp}\bigl(\mp{1\over 2}\sum(\a _i+\aa _i)\bigr)
\prod _{i<j}2\text{sh}{1\over 2}(\a _i-\a _j)
\prod _{i<j}2\text{sh}{1\over 2}(\aa _i-\aa _j)
\prod _{i,j}2\text{ch}{1\over 2}(\a _i-\aa _j)\\
&\times
\frak{F}^{\gamma _1}_{n\ \xi _1}
(\a _1,\cdots,\a _{n-1}|\b_1,\cdots,\b_{2n})\otimes
\frak{F}^{\gamma _2}_{n\ \xi _2}
(\aa _1,\cdots,\aa _{n-1}|\b_1,\cdots,\b_{2n})   \tag {18}
\endalign $$
where
$$
d_{\xi _1 \xi _2}=\pi d_{\xi _1}d_{\xi _2},\qquad
{\zeta}_{\xi _1\xi _2}(\b) = \text{cth}{1\over 2}(\b){\zeta}_{\xi _1}(\b)
{\zeta}_{\xi _2}(\b)
$$
Let us prove that for $n>1$ the functions (17) do satisfy the
residue condition (the case $n=1$ is special). To do that it is
sufficient to realize that under the integral due to antisymmetry
of $\frak{F}$ we can perform a replacement:
$$\align
&\prod _{i<j}\text{sh}{1\over 2}(\a _i-\a _j)
\prod _{i<j}\text{sh}{1\over 2}(\aa _i-\aa _j)
\prod _{1\le i}\prod _{1\le j}\text{ch}{1\over 2}(\a _i-\aa _j)\to
2^{-2(n-2)}e^{(n-2)\b _{2n-1}}
\\
&\times
\bigl\{
(-1)^n(n-1)^2
\text{exp}{1\over 2}\bigl(-(n-2)(\a _1+\aa _1)\bigr)
\text{ch}{1\over 2}(\a _1-\aa _1)
\prod\limits _{i=2}^{n-1}
\text{ch}(\a _i-\b _{2n-1})\\ &\times
\prod\limits _{i=2}^{n-1}
\text{ch}(\aa _i-\b _{2n-1})
\prod _{2\le i<j}\text{sh}{1\over 2}(\a _i-\a _j)
\prod _{2\le i<j}\text{sh}{1\over 2}(\aa _i-\aa _j)
\prod _{2\le i}\prod _{2\le j}\text{ch}{1\over 2}(\a _i-\aa _j)+ \\ \tag {19}
\\&+
{(n-1)(n-2)\over 2}
\text{exp}{1\over 2}\bigl(-(n-2)(\a _1+\a_2)\bigr)
\text{sh}{1\over 2}(\a _1-\a _2)
\prod\limits _{i=3}^{n-1}
\text{ch}(\a _i-\b _{2n-1})\\ &\times
\prod\limits _{i=1}^{n-1}
\text{ch}(\aa _i-\b _{2n-1})
\prod _{3\le i<j}\text{sh}{1\over 2}(\a _i-\a _j)
\prod _{1\le i<j}\text{sh}{1\over 2}(\aa _i-\aa _j)
\prod _{3\le i}\prod _{1\le j}\text{ch}{1\over 2}(\a _i-\aa _j)+\\&{ }\\
\endalign $$

$$\align
&+
{(n-1)(n-2)\over 2}
\text{exp}{1\over 2}\bigl(-(n-2)(\aa _1+\aa _2)\bigr)
\text{sh}{1\over 2}(\aa _1-\aa _2)
\prod\limits _{i=3}^{n-1}
\text{ch}(\aa _i-\b _{2n-1})\\ &\times
\prod\limits _{i=1}^{n-1}
\text{ch}(\a _i-\b _{2n-1})
\prod _{3\le i<j}\text{sh}{1\over 2}(\aa _i-\aa _j)
\prod _{1\le i<j}\text{sh}{1\over 2}(\a _i-\a _j)
\prod _{3\le i}\prod _{1\le j}\text{ch}{1\over 2}(\aa _i-\a _j)\bigr\}
\endalign $$
This formula looks quite terrible, but it has a simple origin:
the LHS is proportional to Vandermonde determinant composed of
$e^{\a _i}$ and $-e^{\aa _i}$, so, to get (19) we added in this
determinant to $k$-th (for $k\ge 3$) row the
$(k-2)$-th one multiplied by $e^{2\b_{2n-1}}$, and then decomposed with respect
to first two rows.  Notice now that we are interested in the second order pole
at $\b _{2n}=\b _{2n-1}+\pi i$ of the integral from (18) since
$\zeta _{\xi _1,\xi _2}(\b _{2n}-\b _{2n-1})$ has zero at this point. But the
contributions from the last two terms in (19) do not have such singularity.
Consider, for example the last term.  The integral with respect to $\aa$'s will
produce first order pole, but the integral with respect to $\a$'s is regular
because it contains $\prod\limits _{i=1}^{n-1}
\text{ch}(\a _i-\b _{2n-1}) $ (see (11)).
Thus the only interesting term is the
first one from (19).
In this term we can replace $\text{ch}{1\over 2}(\a _1-\aa
_1)$ by $$
  (1-ie^{\a _1-\b_{2n-1}})(1+ie^{-\aa _1+\b_{2n-1}})\bigr] 
$$ and then use the formulae (11). That proves the formula (16) for $n\ge 2$.
Let us mention one more important property of these form factors:  they satisfy
the relations:  $$ \bigl(\sum\limits_{j=1}^{2n} e^{-\b _j}\bigr)
f^{\gamma _1\gamma _2}_{+\ \xi _1\xi _2} (\b_1,\cdots,\b _{2n})=
\bigl(\sum\limits_{j=1}^{2n} e^{\b _j}\bigr)
f^{\gamma _1\gamma _2}_{-\ \xi _1\xi _2} (\b_1,\cdots,\b _{2n}) \tag {20}
$$
these relations are proven using the formula (10).
As usual,
to understand what kind of operators do these form factors
correspond to we have to investigate better the two-particle ones.
But before doing that we have to explain what kind
of theory we deal with. The point is that the S-matrix in question
describes many different models.

\head
5. Different models described by tensor product S-matrix
\endhead

The simplest way is to understand the S-matrix as it is, and to
introduce the scalar product in the space of states such that
it respects the Hermitian conjugation of S-matrix:
$$ (S_{\xi _1 \xi _2}(\b))^* = S_{\xi _1 \xi _2}(-\b )$$
Then the symmetry of the model is $U(1)\otimes U(1)$,
two-particle form factors are given by
$$ \align
&f^{\gamma _1\gamma _2}_{\pm\ \xi _1\xi _2} (\b_1,\b _{2})=
d_{\xi _1 \xi _2}
\text{exp}\bigl(\pm{1\over 2}\sum\b _j\bigr)
 \frac {{\zeta}_{\xi _1\xi _2}(\b _1-\b _2)}
{ \text{sh}{\pi \over \xi _1}(\b _2-\b _1-\pi i)
\text{sh}{\pi \over \xi _2}(\b _2-\b _1-\pi i)} \\
&\times
\bigl[ e^{\gamma _1{\pi\over 2\xi _1}
(\b _2-\b _1-\pi i)} \bigl(e_{1,+}\otimes e_{2,-}\bigr) +
e^{-\gamma _1{\pi\over 2\xi _1}
(\b _2-\b _1-\pi i)} \bigl(e_{1,-}\otimes e_{2,+}\bigr)\bigr]\otimes \\
&\otimes
\bigl[ e^{\gamma _2{\pi\over 2\xi _2}
(\b _2-\b _1-\pi i)} \bigl(e_{1,+}\otimes e_{2,-}\bigr) +
e^{-\gamma _2{\pi\over 2\xi _2}
(\b _2-\b _1-\pi i)} \bigl(e_{1,-}\otimes e_{2,+}\bigr)\bigr]
\endalign $$
{}From these expressions an from the formula (20)
one realizes that the form factors of energy-momentum
tensor $T_{\sigma _1 \sigma _2}$ and two $U(1)$ currents
$J_{\sigma}^L$,  $J_{\sigma}^R$ (we use light-cone components,
$\sigma =\pm$) are given by
$$\align
&f_{++} (\b_1,\cdots,\b _{2n})=
\bigl(\sum\limits_{j=1}^{2n} e^{\b _j}\bigr)\sum _{\gamma _1\gamma _2}
f^{\gamma _1\gamma _2}_{+\ \xi _1\xi _2} (\b_1,\cdots,\b _{2n}),\\
&f_{--} (\b_1,\cdots,\b _{2n})=
\bigl(\sum\limits_{j=1}^{2n} e^{-\b _j}\bigr)\sum _{\gamma _1\gamma _2}
f^{\gamma _1\gamma _2}_{-\ \xi _1\xi _2} (\b_1,\cdots,\b _{2n}),\\
&f_{+-} (\b_1,\cdots,\b _{2n})=\\&=
\bigl(\sum\limits_{j=1}^{2n} e^{-\b _j}\bigr)\sum _{\gamma _1\gamma _2}
f^{\gamma _1\gamma _2}_{+\ \xi _1\xi _2} (\b_1,\cdots,\b _{2n})=
\bigl(\sum\limits_{j=1}^{2n} e^{\b _j}\bigr)\sum _{\gamma _1\gamma _2}
f^{\gamma _1\gamma _2}_{+\ \xi _1\xi _2} (\b_1,\cdots,\b _{2n}),\\
&f^L_{\pm} (\b_1,\cdots,\b _{2n})=
\sum _{\gamma _1\gamma _2}
(-)^{\gamma _1} f^{\gamma _1\gamma _2}_{\pm\ \xi _1\xi _2}
(\b_1,\cdots,\b _{2n}),\\
&f^R_{\pm} (\b_1,\cdots,\b _{2n})=
\sum _{\gamma _1\gamma _2}
(-)^{\gamma _2} f^{\gamma _1\gamma _2}_{\pm\ \xi _1\xi _2} (\b_1,\cdots,\b
_{2n})
\endalign $$
The important limit of this model is $\xi _1,\xi _2 \to \infty$.
In this limit we get the PCF.

Now we can consider two levels of restriction of the model.
First, let us introduce the modified energy-momentum tensor
and modified left current as
$$\align
&f_{++}^+ (\b_1,\cdots,\b _{2n})=
\bigl(\sum\limits_{j=1}^{2n} e^{\b _j}\bigr)\sum _{\gamma _1}
f^{\gamma _1\ +}_{+\ \xi _1\xi _2} (\b_1,\cdots,\b _{2n}),\\
&f_{--}^+ (\b_1,\cdots,\b _{2n})=
\bigl(\sum\limits_{j=1}^{2n} e^{-\b _j}\bigr)\sum _{\gamma _1}
f^{\gamma _1\ +}_{-\ \xi _1\xi _2} (\b_1,\cdots,\b _{2n}),
\\
&f_{+-}^+ (\b_1,\cdots,\b _{2n})=\\&=
\bigl(\sum\limits_{j=1}^{2n} e^{-\b _j}\bigr)\sum _{\gamma _1}
f^{\gamma _1\ +}_{+\ \xi _1\xi _2} (\b_1,\cdots,\b _{2n})=
\bigl(\sum\limits_{j=1}^{2n} e^{\b _j}\bigr)\sum _{\gamma _1}
f^{\gamma _1\ +}_{+\ \xi _1\xi _2} (\b_1,\cdots,\b _{2n}),\\
&f^{L,+}_{\pm} (\b_1,\cdots,\b _{2n})=
\sum _{\gamma _1}
(-)^{\gamma _1} f^{\gamma _1\ +}_{\pm\ \xi _1\xi _2}
(\b_1,\cdots,\b _{2n}),\\
\endalign $$
These operators are not selfajoint in SG Hermitian
structure, but they are invariant under the action of
one $U_{q_2}(sl(2))$. That is why for rational ${\xi _2\over \pi}$
the correlations of these operators among themselves contain
only RSOS restricted with respect to $U_{q_2}(sl(2))$ states.
For  ${\xi _2\over \pi}=k+2, k\in \Bbb{Z}$ the model with restricted
set of operators and truncated space can be equipped with
positively defined scalar product.  It is no wonder
that  $J_{\sigma}^R$ is lost in the restricted model: right $U(1)$
is broken.
The important limit of the restricted model is $\xi _1 \to \infty$
for given $k$. In this limit we get perturbations of WZNW on level $k$.

It is instructive to recover SG model itself. It should
coincide with the restricted model for $\xi _1=\xi$,
$\xi _2=3\pi$. In that case in the truncated space of states
right degrees of freedom are frozen and the
restriction of $S_{\xi _2}(\b)$ is just $-1$ (Ising S-matrix).
Consider, for example, the formula
$$ \align
&f_{++}^+ (\b_1,\cdots,\b _{2n})=(2\pi)^n
\bigl(\sum\limits_{j=1}^{2n} e^{\b _j}\bigr)\text{exp}\bigl(\pm{1\over 2}\sum\b
_j\bigr)
\prod _{i<j} \text{cth}{1\over 2}(\b _i-\b _j)
\\&\times \sum _{\gamma }
\sum\limits _{N_{+}=K\cup L}
\f^{\gamma }_{\xi ,\ K} (\b_1,\cdots,\b _{2n})\otimes
\f^{+}_{3\pi,\ L} (\b_1,\cdots,\b _{2n})
\endalign $$
It can be shown that the restriction of
$$\f^{+}_{3\pi,\ L} (\b_1,\cdots,\b _{2n}) $$ differs from zero only
for $L=\{n-2j,\ j=1,\cdots ,n-1\}$. But in the latter case the restriction
coincides (14) with the form factor of Ising disorder operator which is
given by
$$ \left( {1\over 2\pi}\right)^n\prod _{i<j} \text{th}{1\over 2}(\b _i-\b _j)$$
this expression cancels in (17) and we recover SG energy-momentum tensor
form factors (13).

At last if we consider another modification of the
energy-momentum tensor with form factors
$$\align
&f_{++}^{++} (\b_1,\cdots,\b _{2n})=
\bigl(\sum\limits_{j=1}^{2n} e^{\b _j}\bigr)
f^{+ +}_{+\ \xi _1\xi _2} (\b_1,\cdots,\b _{2n}),\\
&f_{--}^{++} (\b_1,\cdots,\b _{2n})=
\bigl(\sum\limits_{j=1}^{2n} e^{-\b _j}\bigr)
f^{+ +}_{-\ \xi _1\xi _2} (\b_1,\cdots,\b _{2n}),\\
&f_{+-}^{++} (\b_1,\cdots,\b _{2n})=\\&=
\bigl(\sum\limits_{j=1}^{2n} e^{-\b _j}\bigr)
f^{++}_{+\ \xi _1\xi _2} (\b_1,\cdots,\b _{2n})=
\bigl(\sum\limits_{j=1}^{2n} e^{\b _j}\bigr)
f^{++}_{+\ \xi _1\xi _2} (\b_1,\cdots,\b _{2n}),\\
\endalign $$
then for ${\xi _1\over \pi},\ {\xi _2\over \pi} $ integers
we can perform restriction which will give the perturbations
of coset models discussed in [11].

It should be said that we restricted ourselves in this paper with
consideration of repulsive SG coupling constants ($\xi \ge \pi$).
The consideration of attractive case is also important,
after restriction we can find quite unexpected models.

\head
6. The mathematical structure of the solution.
\endhead

The S-matrix
$ - S^{SG}_{\xi _1}(\b)\otimes S^{SG}_{\xi _2}(\b) $
is constructed in amusing way: as if we have two types of
particles confined together. The formula for form factors
$$ \align
&f^{\gamma _1\gamma _2}_{\pm\ \xi _1\xi _2} (\b_1,\cdots,\b _{2n})=
\text{exp}\bigl(\pm{1\over 2}\sum\b _j\bigr)
\prod _{i<j} \text{cth}{1\over 2}(\b _i-\b _j)
\\&\times \sum\limits _{N_{\pm}=K\cup L}
\f^{\gamma _1}_{\xi _1,\ K} (\b_1,\cdots,\b _{2n})\otimes
\f^{\gamma _2}_{\xi _2,\ L} (\b_1,\cdots,\b _{2n}) \tag {21}
\endalign $$
shows that these two types of particles "interact" in quite
interesting fashion. In this section we shall explain the
most attractive features of this "interaction".

When considering the solutions to Riemann-Hilbert problem
we explained that the sets $K$ which count different
solutions play the same role as choice of different contours in
integral formulae for the solutions
of KZ equations. From the algebraic
point of view these data count different
passes composed of intermediate Verma modules
for the products of vertex operators whose
vacuum expectations provide the solutions to KZ and
deformed KZ equations. This interaction through
passes looks quite interesting, it is similar to
combining together holomorphic
and antiholomorphic conformal blocks in CFT.

However, we suppose the following circumstance to be of the
main importance. We deal with $k=0$ equations, in that case the
contours in question are quite special, as it has been mentioned in
Section 2. Let us explain this point in more details, but
before doing that we have to provide some more information.

Let us concentrate on one
SG S-matrix with coupling constant $\xi$.
First, we shall give more explicit description
of the vector
$\frak{F}^{\gamma}_n
(\a _1,\cdots,\a _{n-1}|\b_1,\cdots,\b_{2n})$.

We introduce the notation $B=\{1,2,\cdots,2n\}$.
To every $T\subset B:\ \# T=n$ vector $w_T (\b_1,\cdots,\b _{2n-1},\b_{2n})$
from the tensor product
of isotopic spaces is related such that
$$ w_T(\b_1,\cdots,\b _i,\b
_{i+1},\cdots,\b_{2n}) \widehat{S}_{\xi}(\b _i-\b _{i+1})
=w_{T\backslash i}(\b_1,\cdots,\b _{i+1},\b _i,
\cdots,\b_{2n}), $$
where
$$T\backslash i=\left\{ \matrix &T\qquad &\text{if}\ i,i+1\in T\  \text{or}
\ i,i+1 \notin T\\&T\circ {i}\circ \{i+1\}\qquad &\text{if}
\ i\in T,\ i+1\notin  T\ \text{or}
\ i+1 \in T,\ i \notin T \endmatrix \right. $$
where $A\circ B=(A\cup B)\backslash (A\cap B)$.
These notation are different from those used in [1], but
one can easily understand the relation.
It would simplify a lot further formulae if we introduce
notations
$$v_T^{\gamma} (\b_1,\cdots,\b_{2n}) =
{ \text{exp}{\gamma\pi\over 2\xi}
\bigl(\sum _{j\in T}\b _j-\sum _{j\notin T}\b _j \bigr)
\over \prod\limits _{i\in T,j\notin T}\text{sh}{\pi\over\xi}
(\b _i-\b _j)}
w_T (\b_1,\cdots,\b_{2n}) $$
We have [1]:
$$ \align
&\frak{F}^{\gamma}_n
(\a _1,\cdots,\a _{n-1}|\b_1,\cdots,\b_{2n})=\\&=
\sum\limits _{T\in B,\# T=n}
\Delta ^T (\a _1,\cdots,\a _{n-1}|\b_1,\cdots,\b_{2n})
v_T^{\gamma} (\b_1,\cdots,\b_{2n}) \endalign $$
where  $\Delta ^T $ is $(n-1)\times (n-1) $ matrix with matrix elements:
$$
A_{i,j}^T=A_i^T(\a _j|\b_1,\cdots,\b_{2n}), $$
where
$$ \align &
A_i^T(\a|\b_1,\cdots,\b_{2n})=\text{exp}(-{\pi\over \xi}
((n-2)\a +\sum \b _p)) \\&\times
\left\{\prod\limits _{q\in T}(e^{{2\pi\over \xi}\a}-q^{1\over 2}
e^{{2\pi\over \xi}\b _q})\sum\limits _{k=0}^{i-1}(1-q^{i-k})
e^{{2\pi\over \xi}(i-k-1)\a}q^{k\over 2}\sigma _{k,\ B\backslash T} +\right.\\&
+q^i
\left.\prod\limits _{q\in B\backslash T}(e^{{2\pi\over \xi}\a}-q^{-{1\over 2}}
e^{{2\pi\over \xi}\b _q})\sum\limits _{k=0}^{i-1}(1-q^{i-k})
e^{{2\pi\over \xi}(i-k-1)\a}q^{k\over 2}\sigma _{k, T} \right\}
\endalign $$
with $  \sigma _{k, T}$,  $\sigma _{k,\ B\backslash T}$ are
elementary symmetric polynomials of degree $k$ with arguments
$\text{exp}{2\pi\over \xi}(\sum \b _p),\ p\in T $ and
$\text{exp}{2\pi\over \xi}(\sum \b _p),\ p\notin T $ respectively.
We remind also that $q= \text{exp}{2\pi ^2 i\over \xi}$.

These formulae provide that the functions
$\f^{\gamma}_{\xi,\ K} (\b_1,\cdots,\b _{2n})$ can be written in the form:
$$\align
&\f^{\gamma}_{\xi,\ K} (\b_1,\cdots,\b _{2n})=\\&=
\sum\limits _{T\in B,\# T=n}
\text{det}\left|\int \prod _p \varphi(\a-\b _p)
A_i^T(\a|\b_1,\cdots,\b_{2n}) \text{exp}(k_j\a)d\a \right|
v_T^{\gamma} (\b_1,\cdots,\b_{2n}) \tag {22}\endalign $$
where $K=\{k_1,\cdots,k_{n-1}\}$, $k_1<k_2<\cdots <k_{n-1}$.

To rewrite this answer in more beautiful way let us introduce
two vector spaces: the space $H$ of dimension $2n-2$ with basis
$A_i,B_i \ i=1,\cdots, n-1$, and the space $V$ of dimension $n-1$ with
basis $Z_i , \ i=1,\cdots, n-1$. Now for every $T\subset B,\ \#T=n$ we
introduce two forms
$$\omega ^{\pm}(T)=\omega _{i,j}(T)A_i\wedge Z_j\pm
\tilde{\omega} _{i,j}(T)B_i\wedge Z_j   \tag {23} $$
where
$$ \align &
\omega _{i,j}(T)=\text{exp}(-{1\over 2}\sum \b _p)
\bigl\{
\sum\limits _{i=1}^{n-1} \sum\limits _{j=1}^{n-1}
\sigma _{2n-2j}(e^{\b _1},\cdots ,e^{\b _{2n}})\\&\times
\int  \prod _p \varphi(\a-\b _p)
A_i^T(\a|\b_1,\cdots,\b_{2n}) \text{exp}((2p-n)\a)d\a \bigr\}
\\{ }\\
&
\tilde{\omega} _{i,j}(T)= \text{exp}(-{1\over 2}\sum \b _p)\bigl\{
\sum\limits _{i=1}^{n-1} \sum\limits _{j=1}^{n-1}
\sum\limits _{l=1}^{j}
\sigma _{2n-2l+1}(e^{\b _1},\cdots ,e^{\b _{2n}}) \\&\times
\int  \prod _p \varphi(\a-\b _p)
A_i^T(\a|\b_1,\cdots,\b_{2n}) \text{exp}((2l-n-1)\a)d\a
\bigr\}
\endalign $$
Then different solutions (22) can be found as coefficients
in decomposition with respect to $$({\pm})^q A_{j_1}\wedge\cdots \wedge A_{j_q}
\wedge B_{k_1}\wedge\cdots \wedge B_{k_{n-q-1}}\tag {24}$$
of the expression
$$\sum\limits _{T\in B,\# T=n}\wedge ^{(n-1)} (\omega ^{\pm}(T))
v_T^{\gamma} (\b_1,\cdots,\b_{2n}) \tag {25}$$
We can think of this decomposition as of one with
respect to Grassmanian of $(n-1)$-dimensional subspaces of $H$.

The "interaction" in the formula for
the form factor $f^{\gamma _1\gamma _2}_{+\ \xi _1\xi _2}$ (21)
can be understood as follows. Take two copies of the spase $V$
(with basises $Z_i$, $W_i$). Then the form factors are defined
by the inner product of forms $\wedge ^{(n-1)}\omega ^+_{\xi _1}$
and $\wedge ^{(n-1)}\omega ^-_{\xi _2}$ as follows
$$\align
d_{\xi _1 \xi _2}^n
\text{exp}\bigl({n\over 2}\sum\b _j\bigr) &
\prod _{i<j} {\zeta}_{\xi _1\xi _2}(\b _i-\b _j)
\bigl(\prod\limits _{p=1}^{2n-1}
\sigma _{1}(e^{\b _1},\cdots ,e^{\b _{2n}}) \bigr)^{-2}\times\\
\sum\limits _{T_1\in B,\# T=n}
\sum\limits _{T_2\in B,\# T=n}
&\wedge ^{(n-1)}(\omega _{\xi _1,\ i,j}(T_1)A_i\wedge Z_j +
\tilde{\omega} _{\xi _1,\ i,j}(T_1)B_i\wedge Z_j)\wedge\\
&\wedge ^{(n-1)}(\omega _{\xi _2,\ i,j}(T_2)A_i\wedge W_j -
\tilde{\omega} _{\xi _2,\ i,j}(T_2)B_i\wedge W_j)
\\ &\times
v_{\xi _1,T_1}^{\gamma _1} (\b_1,\cdots,\b_{2n})\otimes
v_{\xi _2,T_2}^{\gamma _2} (\b_1,\cdots,\b_{2n}) = \\
=f^{\gamma _1\gamma _2}_{\pm\ \xi _1\xi _2} (\b_1,\cdots,\b _{2n})&
\\ \times A_1\wedge \cdots\wedge A_{n-1}\wedge
    B&_1\wedge \cdots\wedge B_{n-1}\wedge
    Z_1\wedge \cdots\wedge Z_{n-1}\wedge
    W_1\wedge \cdots\wedge W_{n-1}
\endalign $$
Certainly, the formula (25) presents rather formal way of rewriting (22),
but it is not an empty exercise since the form $\omega (T)$
independently has interesting meaning. It has been said in Section 2
that the expression
$$\align &\text{exp}(-{1\over 2}\sum \b _p)\sigma _{n-k}
(e^{\b _1},\cdots ,e^{\b _{2n}})\\ &\times
\int  \prod _p \varphi(\a-\b _p)
A_i^T(\a|\b_1,\cdots,\b_{2n}) \text{exp}(k\a)d\a
\endalign $$
can be considered as quantum deformation of the period
of special second type hyperelliptic differential
$\zeta^T _i$
over the contour around two $c_k$ around two
branching points $\b _{n+k}$ and
$\b _{n+k+1}$. The canonical choice of homology basis on the
surface is the following:
$$a_i=c_{2i-n},\ b_i =c_{-n+1}+c_{-n+3}+\cdots +c_{-n+2i-1}$$
for $i=1,\cdots ,(n-1)$, genus of surface equals $n-1$.
{}From that point of view it is natural
to identify $A_i, B_i$ with basic vectors of the lattice of
periods and the vectors $Z_i$ with differentials along
Jacobian. Then the form $\omega (T) $ has nice meaning [21].
We shall explain this point in some more details in the next
section. The combination of $\omega ^+$ with $\omega ^-$
reminds one more time combining holomorphic and antiholomorphic
pieces in CFT: the $b$-cycles are imaginary ones.

\head
7. Remarks on the classical limit.
\endhead

Let us discuss the perturbation of WZNW model on level $k$.
In that case we have to put $\xi _1=\infty$,
$\xi _2=k+2$, to fix $\gamma _2$  (say $\gamma _2 =+$ )
and to perform restriction. Let us consider, for example,
the form factors of one component of energy-momentum tensor:
$$ \align
&f_{++} (\b_1,\cdots,\b _{2n})= (2\pi)^n
\text{exp}\bigl({1\over 2}\sum\b _j\bigr)
\bigl(\sum\limits e^{\b _j}\bigr)
\prod _{i<j} \text{cth}{1\over 2}(\b _i-\b _j)
\\&\times \sum\limits _{N_{+}=K\cup L}
\f_K(\b_1,\cdots,\b _{2n})\otimes
\f^{k}_{L} (\b_1,\cdots,\b _{2n})  \tag {26}
\endalign $$
where the notations are used
$$\align
&
\f_K(\b_1,\cdots,\b _{2n})=\text{lim}_{\xi\to\infty}
\f^{\gamma}_{\xi _1,\ K} (\b_1,\cdots,\b _{2n}),\\
&
\f^{k}_{L} (\b_1,\cdots,\b _{2n})=
(\f^{+}_{\pi (k+2),\ L} (\b_1,\cdots,\b _{2n}))_{\text{restricted}}
\tag {27}\endalign $$
Two pieces of (26) looking quite similar have very different
analytical structure. The function
$\f_K(\b_1,\cdots,\b _{2n}) $ is a solution to
Yangian Riemann-Hilbert problem i.e. that with rational in $\b$
S-matrix. That is why it makes perfect sense to consider the following
classical limit for this piece [20]:
\def\la{\lambda}
$$\b _j={2\pi  \la _j \over \hbar} \qquad \hbar \to +0$$
In that limit the asymptotics of
$\f_K(\b_1,\cdots,\b _{2n}) $ is described by
a solution to usual KZ equations on level zero.
On the other hand the S-matrix for
$\f^{k}_{L} (\b_1,\cdots,\b _{2n}) $ depends typically
on $\text{exp}({\beta _j\over k+2})$, and the limit (27)
does not make much sense for this function.
Our dream would be to present the form factor as
functional integral (with special boundary conditions)
of the original action of the theory. The theory has two main features:
it is asymptotically free and it contains WZNW term in action.
These two features should lead to different effects:
the asymptotic freedom should provide reasonable
perturbation theory with respect to Plank constant while
the WZNW term should provide nontrivial nonperturbative
effects. The exact solution (26) shows that these two effects are combined
in rather special way. Notice that the first piece of (26) is
independent of $k$, also it allows the classical limit (27),
$\hbar$ should be identified with Plank constant: it rescales
the rapidities (logarithms of momenta) which makes perfect
sense in asymptotically free situation. The second piece is
the one depending upon $k$, it should be related to
nonperturbative effects due to WZNW term. The formula (26) and
the Grassmanian of the previous section
suggest that by introducing certain fermionic field
we should be able to treat these two pieces independently
and perform the averaging over this fermionic field
(summation over $K,L$) afterwards.
We hope to return to the consideration of functional
integral in feature, but now let us concentrate
on the quasiclassical theory of
$\f_K(\b_1,\cdots,\b _{2n}) $.

What we shall do now is certain extension of the
consideration of papers [20,21]. Let us take the form $ \omega (T) $ from the
previous section for the Yangian case ($\xi=\infty$). Now we calculate the
asymptotics (27) the result being [20]:
$$
\omega^{\pm} (T)=\sum\limits _{i=1}^{n-1}\sum\limits _{j=1}^{n-1}
(\int\limits _{a_j} \zeta^T _i \  A_j\wedge Z_i\pm
\int\limits _{b_j} \zeta^T _i \  B_j\wedge Z_i ) \tag {28}$$
where $a_i,\ b_i$ are basic contours on the hyperelliptic surface
$\tau^2=P(\la)\equiv\prod (\la-\la _p)$
of genus $g=n-1$ , $\zeta^T _i$ is the following second kind
differential:
$$ \align
&\zeta^T _i=
{1\over \sqrt{P(\la)}}
 \\&\times
\left\{
\prod\limits_{p\in T}
(\la-\la _p)
\left[
{d\over d\la}
{\prod _{p\notin T} (\la-\la _p)\over \la^{n-i}}
\right] _+ +
\prod\limits
_{p\notin T} (\la-\la _p)
\left[{d\over d\la}
{\prod _{p\in T} (\la-\la _p)\over \la^{n-i}}
\right]_+
\right\}
\endalign  $$
where $[]_+$ means that only polynomial part of the expression in
brackets is taken.
When substituting $\omega^{\pm} (T) $ into (25) we can perform
certain transformations. Little modification of formulae from [20]
provides that (25) can be rewritten in the limit as
$${\hbar}^{3n\over 4}
C^{-3}\sum\limits _{T\in B,\# T=n}\wedge ^{(n-1)} (\hat{\omega}^{\pm} (T))
\ \teet T 0 ^4 E_T$$
where $E_T$ is basic vector in the tensor product of isotopic
spaces. We introduced Riemann theta-function taking
$a_i,\ b_i$ for the canonical basis, $\eta _T$ is
even nonsingular  (such that $\teet T 0 \ne 0$)
half-period related to the subset of
branching points defined by $T$ [22],
$$C=\prod _{i<j}(\la _i-\la _j)^{\frac 1 4}\Delta , \tag {29}$$  $\Delta$ is
given by the determinant
$$\Delta=\DET \left|\int _{a_i}
{\la^{j-1}\over\sqrt{P(\la)}}\right|_{(n-1)\times
(n-1)}$$ The form $\hat{\omega}^{\pm} (T)$ is rewritten as $$
\tilde{\omega} (T)=\sum\limits _{i=1}^{n-1}\sum\limits _{j=1}^{n-1} \bigl(
\partial _i\partial _j\text{log}\teet {T}{0} A_i\wedge dz_j \pm
\sum\limits _{l=1}^{n-1}
(\Omega _{i,l}\partial _l\partial _j\text{log}\teet {T}{0}
+2\pi i\delta _{i,j})
B_l\wedge dz_j
\bigr) \tag {30}
$$
where we replaced quite formally $Z_i$ by differential along Jacobian
$dz_i$.

Let us consider the form on the Jacobian which interpolates
between (30) for different $T$:
$$
\hat{\omega}^{\pm} (z)=\sum\limits _{i=1}^{n-1}\sum\limits _{j=1}^{n-1} \bigl(
\partial _i\partial _j\text{log}\theta (z) A_i\wedge dz_j\pm
\sum\limits _{l=1}^{n-1}
(\Omega _{i,l}\partial _l\partial _j\text{log}\theta (z)
+2\pi i\delta _{i,j})
B_l\wedge dz_j
\bigr)  $$
where $z$ varies over Jacobian: $z\in\Bbb{C}^{n-1}/ \Bbb{Z}^{n-1}\times
\Omega\Bbb{Z}^{n-1}$
The mathematical meaning of this form can be explained as
averaging of the simplectic form induced on Jacobian
after embedding into projective space by means of
theta-functions of second order [21].

Probaaly the best possible situation takes place for PCF.
Here both pieces of the form factor allow classical limit,
and the form factors are special values (at even non-singular
half-periods) of the form:
$$ {\hbar}^{3n\over 2}C^{-6}
\theta (z)^4\theta (w)^4(\wedge ^{(n-1)}\ \hat{\omega}^{+}(z) ) \wedge
  ( \wedge ^{(n-1)}\ \hat{\omega}^{-}(w) ) \tag {31}$$
It is interesting that all the formulae of this paper can be considered
as deformations of (31).

\head
8. Conclusion.
\endhead
To conclude this paper let us formulate several
problems which, to our mind, are worth investigation.
\item {1.} It would be interesting to generalize the
considerations of this paper to other Lie algebras. The
necessary preliminary information for $SU(N)$ case can be
found in [1].
\item {2.} It is interesting to consider the deformation
of the construction of this paper to the lattice models [23].
In that case different choice of contours in KZ is replaced
not by introducing the exponents under the integrals, as it
happened in the situation of this paper, but to introducing
different theta-functions. The generalization for the
lattice analog of PCF should be not complicated to find,
however, the generalization for lattice version of perturbed
WZNW on level $k$ which coincides with integrable version of 6-vertex model
of spin $2k$ ,$|q|<1$ is not easy to describe because it will
require the knowledge of solutions to elliptic version of
Riemann-Hilbert problem (1) which are not known (the same is needed for
form factors of 8-vertex model).
\item {3.} We suppose that the most important question is that of
understanding the origin of form factor formulae in terms
of functional integral. The results of this paper should be
important in understanding of this problem. Formula (31) should
be crucial for this goal.

{\bf Acknowledgement}
I am grateful to N. Reshetikhin for interesting and
stimulating discussions.

\end

\ref
\key
\by F.A.Smirnov
\jour Int.Jour.Math.Phys.
\vol 4A
\yr 1989
\pages 4213
\endref

\ref
\key
\by T. Eguchi, S.K. Yang
\jour Phys. Lett.
\vol 235B
\yr 1990
\pages 282
\endref

\ref
\key
\by A. LeClair
\jour Phys. Lett.
\vol 230B
\yr 1989
\pages 103
\endref
\end